\documentclass[proceedings, preprint]{rmaa}

\usepackage{paralist}

\usepackage{psfrag,color}

\SetYear{2009}
\SetConfTitle{A Long Walk Through Astronomy}

\title{Barred Galaxy Photometry: Comparing results from the Cananea
  sample with $N$-body simulations} 

\author{
  E. Athanassoula\altaffilmark{1}, 
  D. A. Gadotti\altaffilmark{2},
  L. Carrasco\altaffilmark{3}, 
  A. Bosma\altaffilmark{1}, 
  R. E. de Souza\altaffilmark{4},
  and E. Recillas\altaffilmark{3}}

\altaffiltext{1}{Laboratoire d'Astrophysique de Marseille (LAM), UMR6110,
Observatoire Astronomique de Marseille Provence,
CNRS/Universit\'e de Provence, Technop\^ole de Marseille-Etoile, 
38 rue Fr\'ed\'eric Joliot Curie, 13388 Marseille C\'edex 20, France
(lia@oamp.fr, bosma@oamp.fr).}

\altaffiltext{2}{Max-Planck-Institut f\"ur Astrophysik,
  Karl-Schwarzschild-Str. 1, D-85748 
Garching bei M\"unchen, Germany (dimitri@mpa-garching.mpg.de).}

\altaffiltext{3}{Instituto Nacional de Astrof\'isica, Optica, y
  Electr\'onica, Luis Enrique Erro 1, Tonantzintla, C.P. 72840,
  Puebla, Mexico (carrasco@inaoep.mx, elsare@inaoep.mx).}

\altaffiltext{4}{Departamento de Astronomia, Universidade de S\~ao Paulo,
Rua do Mat\~ao 1226, 05508-090, S\~ao Paulo-SP, Brasil
(ronaldo@astro.iag.usp.br).}  

\shortauthor{Athanassoula et al.}
\shorttitle{Comparing photometric results of real and $N$-body bars}

\listofauthors{E. Athanassoula, D. Gadotti, L. Carrasco, 
A. Bosma, R. E. de Souza, \& E. Recillas}
\indexauthor{Athanassoula, E.}
\indexauthor{Gadotti D.}
\indexauthor{Carrasco, L.}
\indexauthor{Bosma, A.}
\indexauthor{de Souza, R.}
\indexauthor{Recillas, E.}

\abstract{We compare the results of the photometrical analysis of 
barred galaxies with those of a similar analysis from $N$-body
simulations. The photometry is for a sample of nine barred galaxies
observed in the  $J$ and $K_s$ bands with the CANICA near infrared
(NIR) camera at 
the 2.1-m telescope of the Observatorio Astrof\'isico Guillermo Haro
(OAGH) in Cananea, Sonora, Mexico. The comparison includes radial
ellipticity profiles and surface brightness (density for the $N$-body galaxies)
profiles along the bar major and minor axes. We find very good
agreement, arguing that the exchange of angular momentum within the
galaxy plays a determinant role in the evolution of barred galaxies.
}

\resumen{Presentamos los resultados de observaciones de galaxias
barradas y de simulaciones de $N$-cuerpos}

\addkeyword{galaxies: photometry}
\addkeyword{galaxies: structure}
\addkeyword{galaxies: evolution}
\addkeyword{galaxies: general}

\begin{document}
\maketitle

\section{Introduction}
\label{sec:intro}

The majority of disc galaxies are barred \citep[e.g.][]{deVaucouleursdVCBPF91,
  Eskridge00, KnapenSP00, Marinova.Jogee07, MenendezDSSJS07} so that
understanding the formation and evolution of 
bars and their relation to the remaining galactic components is an essential
step towards understanding disc galaxy formation and evolution in
general. We will here briefly describe some work to that avail, 
which compares results from photometric observations 
\citep{Gadotti.ACBSR07} to $N$-body simulations \citep{Atha.Misiriotis02}.

\section{$N$-body simulations}
\label{sec:simulations}

\begin{figure*}
\begin{center}
  \includegraphics[clip=true,width=0.9\columnwidth]{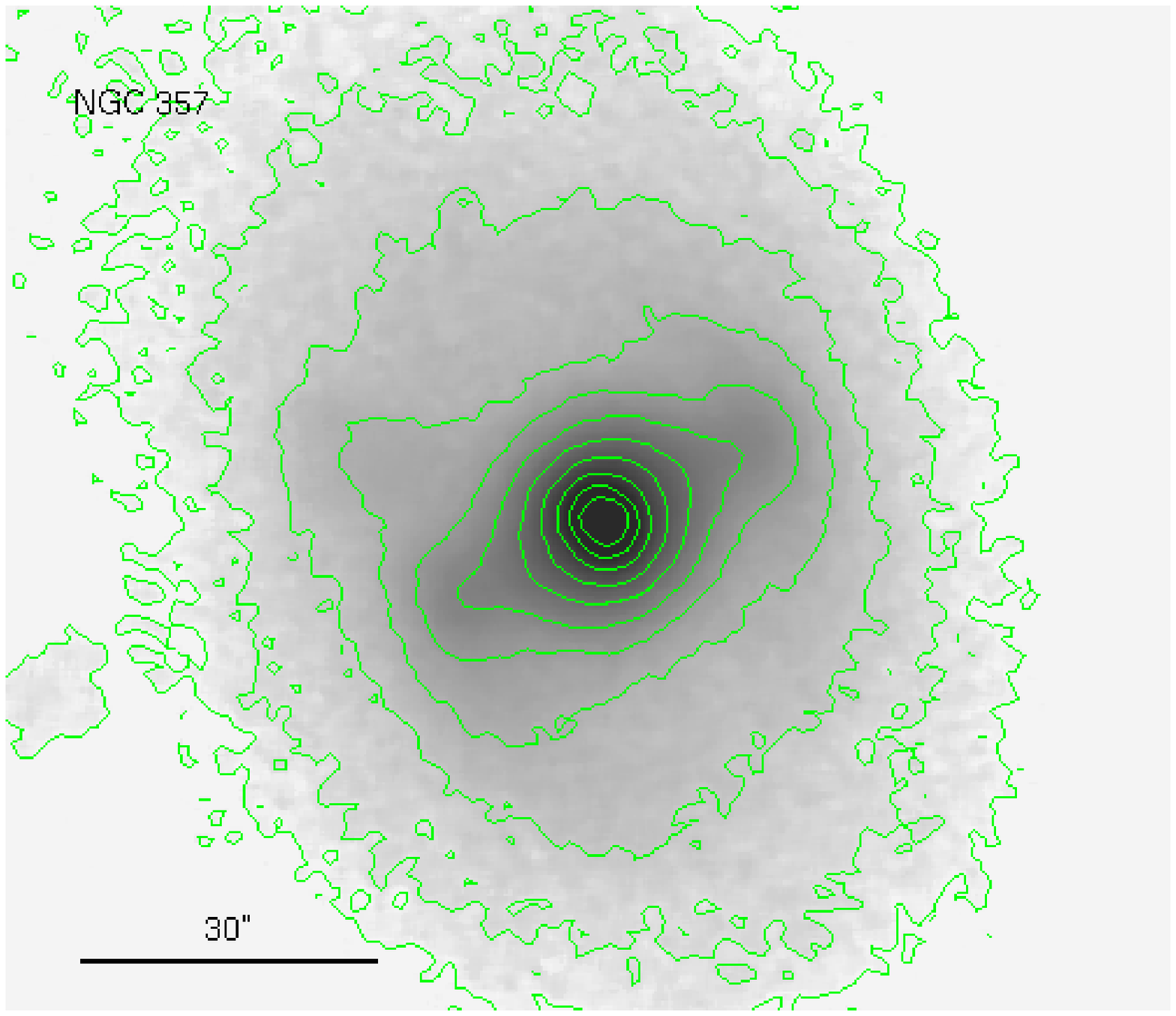}
  \includegraphics[clip=true,width=0.9\columnwidth]{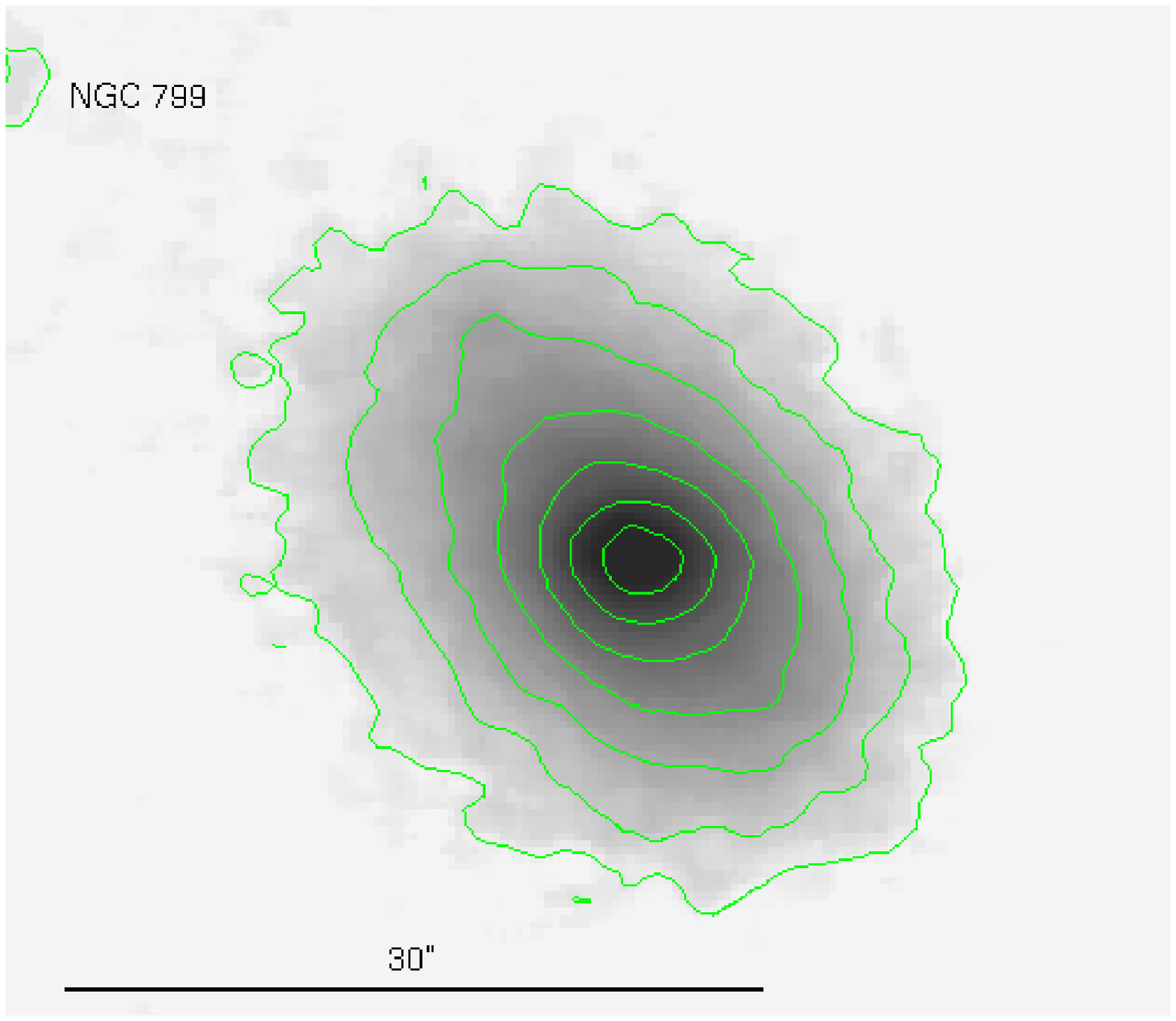}
\end{center}
  \caption{$J$ images of NGC 357 (left) and NGC 799 (right) with a
    few characteristic isophotes superposed.}
  \label{fig:N357_799}
\end{figure*}

Barred galaxies evolve by internal redistribution of angular momentum.
By measuring the frequencies of the orbits in $N$-body simulations,
\citet{Atha02, Atha03} showed that angular momentum is emitted by
near-resonant material in the bar region and 
absorbed by near-resonant material in the outer disc or in the halo. The
strength of the bar increases with the amount of angular exchanged. 
In fact, \citet{Atha03} found a correlation between the angular
momentum acquired by the halo and the bar strength. 

$N$-body simulations show that the
properties of bars which grow in $N$-body galaxies within which a
considerable amount of angular momentum was exchanged are quite
different from the properties of bars grown in $N$-body galaxies 
with little such exchange \citep{Atha.Misiriotis02, Atha03}. This includes the
morphological, the photometrical and the kinematical
properties. $N$-body galaxies with considerable angular momentum
internal redistribution have strong bars which often have ansae and
which, viewed side-on (i.e. edge-on with the line of sight along the
bar minor axis), have a characteristic peanut or `X'-like
shape. On the contrary, $N$-body galaxies within which little angular
momentum has been 
exchanged have weaker bars with no ansae and whose side-on shape is
boxy. These two
types, which we will hereafter refer to as MH and as MD bars, are just
the two extremes of a continuous sequence, and intermediate strength
bars correspond to an exchange of an intermediate amount of angular
momentum. 

Strong differences between the MH and the MD types can be seen in
the ellipticity profiles, the bar shape and the radial projected
density profiles along and perpendicular to the bar. Fig. 4 of 
 \citet{Atha.Misiriotis02} shows that in MH types there is an abrupt
 drop of the ellipticity at a location coinciding roughly with the end
 of the bar, while this feature is lacking from the radial ellipticity
 profiles of MD type $N$-body galaxies. Also the maximum ellipticity
 is much higher for MH types than for MDs. 

MD type bars have isodensities which are well represented by ellipses.
On the contrary MH types have a rather rectangular-like isodensities
particularly in the outer parts of the bar. This can be quantified by
the shape parameter $c$, which gives the shape of a generalised
ellipse

\begin{equation}
(|x|/a)^c + (|y|/b)^c = 1.
\end{equation}

\noindent
Here $a$ and $b$ are the major and minor axes of the generalised
ellipse and $c$ is its shape parameter. 
Ellipses have $c$ = 2, rectangular-like shapes have $c > 2$ and
lozenges $c < 2$ \citep{AthaMWPPLB90}. Fitting the isodensities in the
$N$-body images with generalised ellipses,  \citet{Atha.Misiriotis02}
find that MH cases have considerably larger values of $c$ than MD
cases. Alternatively, it is possible to fit simple ellipses and
measure their departure of the isodensities from the elliptical shape
\citep{jed87}. 
This type of measurement also shows that the shape of the isodensities
in MH type bars departs considerably from ellipses, contrary to MD
types.

Finally,  \citet{Atha.Misiriotis02} made projected radial density
profiles along the minor and the major axes of their $N$-body bars. They
showed that the latter in MH types contributes a flat ledge, dropping
abruptly at a radius which agrees with the end of the bar. On the
other hand the radial projected density in the region of the bar in MD
types shows only a gradual exponential-like decrease. 

\section{Observations}
\label{sec:observations}

\begin{figure*}
\begin{center}
  \includegraphics[clip=true,width=0.5\columnwidth]{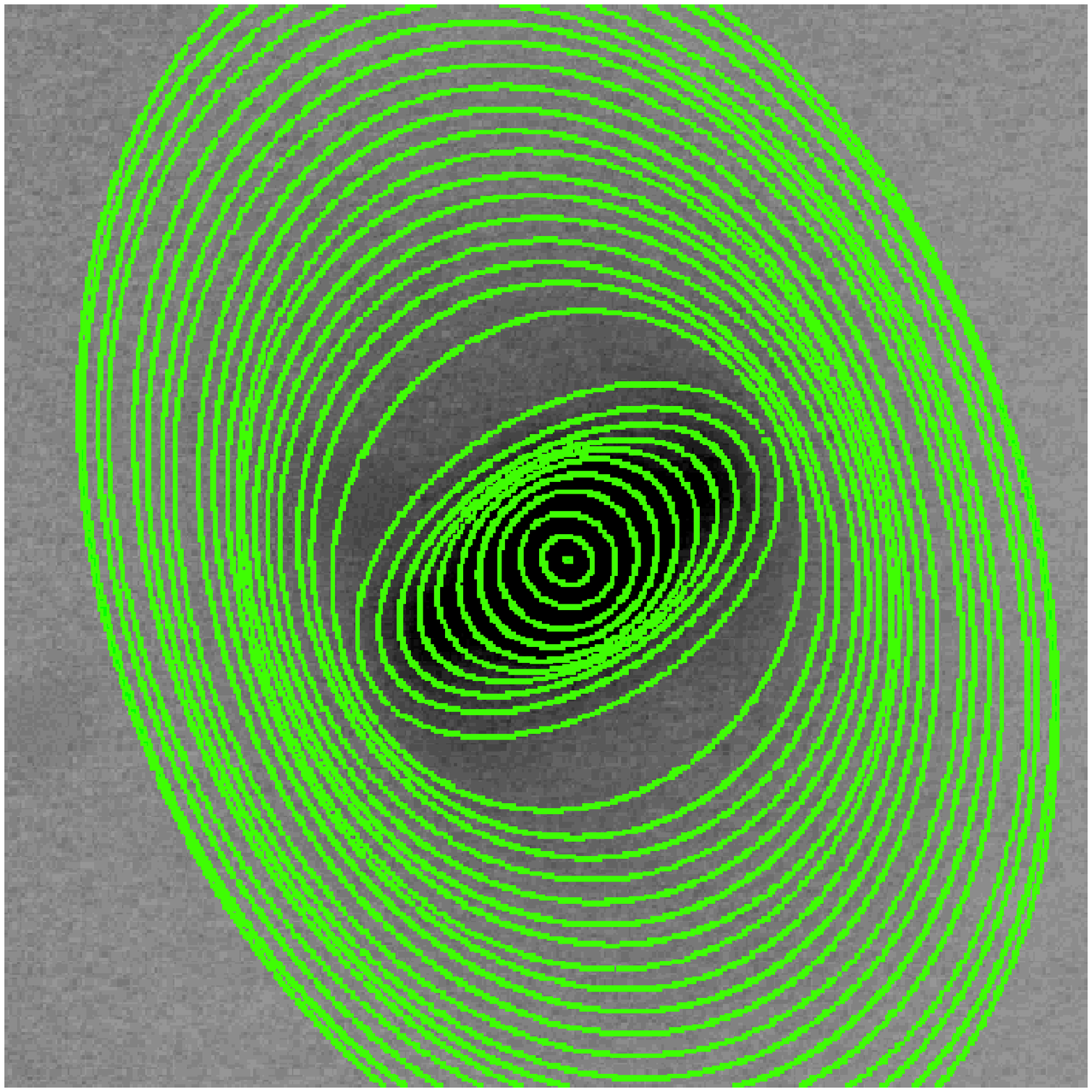}
  \includegraphics[clip=true,width=0.5\columnwidth]{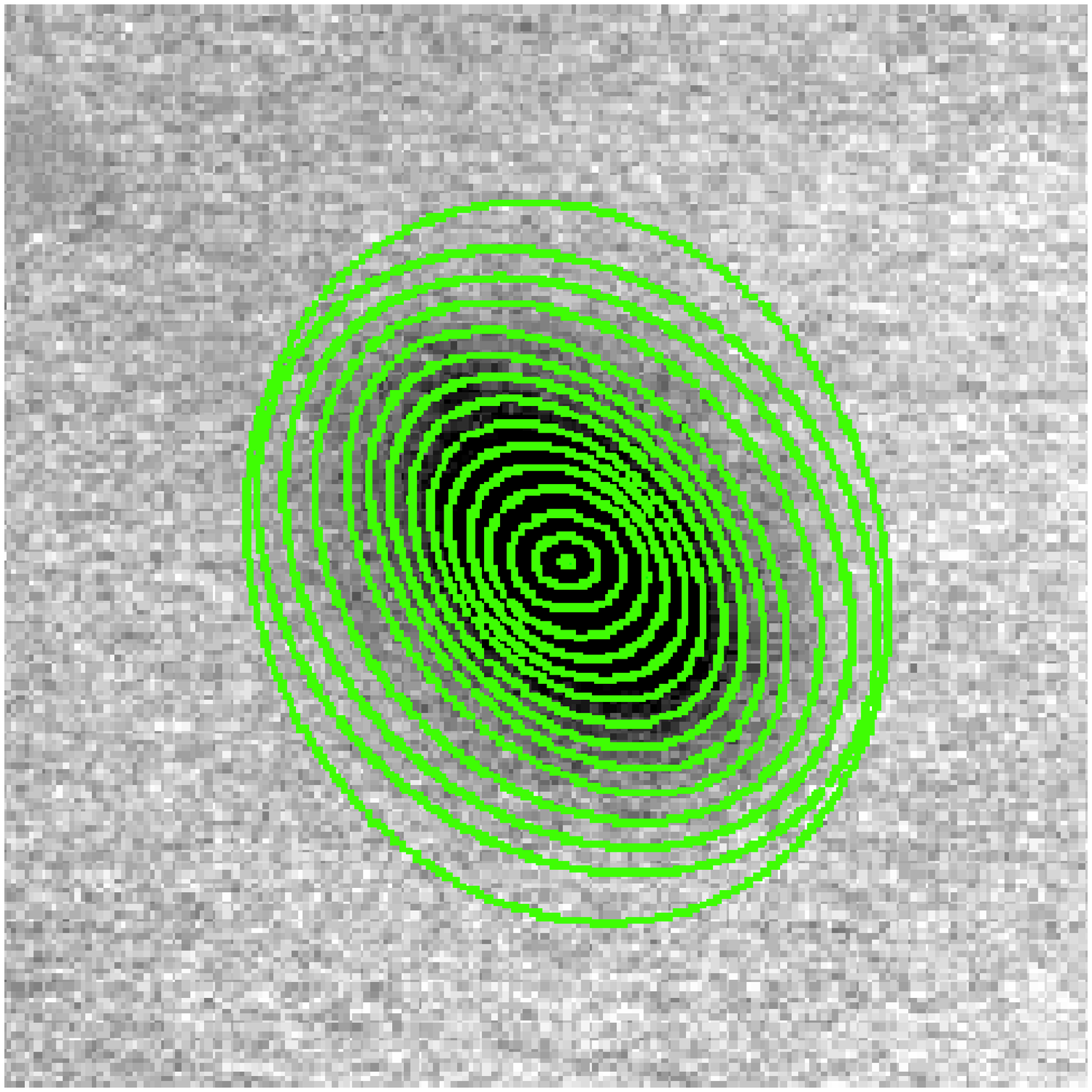}
\end{center}
  \caption{$J$ images of NGC 357 (left) and NGC 799 (right) with, superposed,
    a few of the fitted ellipses.}
  \label{fig:ellipses}
\end{figure*}

We observed a sample of nine barred galaxies in the $J$ and $K_s$
wavelengths using the CANICA near infrared (NIR) camera available at
the 2.1-m telescope of the Observatorio Astrof\'isico Guillermo Haro
(OAGH) in Cananea, Sonora, Mexico. Our aim is to compare real barred
galaxies to $N$-body ones, and that has guided us in the choice of our
sample. This consists of NGC 266, 357, 799, 1211, 1358, 1638, 7080,
7280 and 7743. Our images are deep and thus allowed us to trace the
spiral arms for quite a large azimuthal angle and also to make sure
that the outermost isophotes, which are used for the deprojection, are
not much perturbed by the bar.  

\begin{figure*}
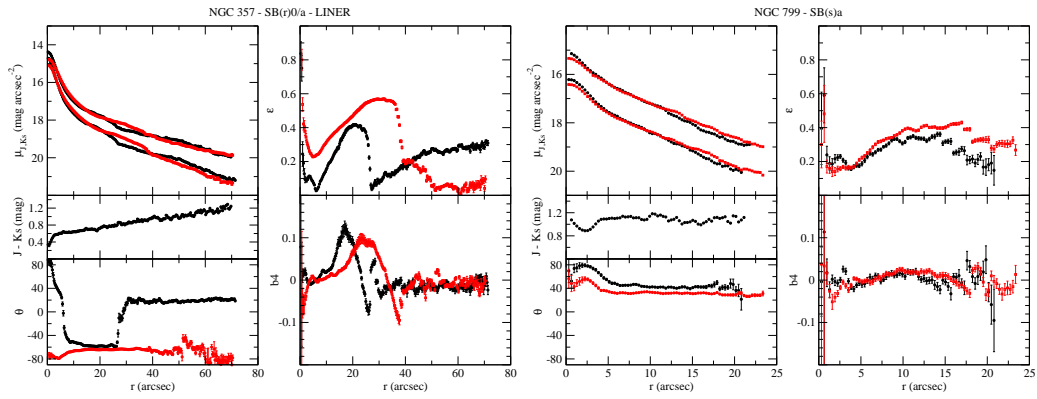

\begin{center}
  \includegraphics[clip=true,width=0.85\columnwidth]{N357.eps}
  \includegraphics[clip=true,width=0.85\columnwidth]{N799.eps}
\end{center}
  \caption{Results of the ellipse fit for NGC 357 (left) and NGC 799
    (right). These include the projected surface brightness in $J$
    and $K_s$ (upper left), the $J-K_s$ colour (center left), the
    ellipse position angle measured from north to east (bottom left),
    ellipticity (upper right) and the $b_4$ Fourier coefficient
    (bottom right). These are given both for the direct and the
    deprojected views, the latter plotted in the online version in
    red.}
  \label{fig:ellipsres}
\end{figure*}

Here we will discuss more specifically two galaxies, NGC 357 and NGC
799, which we will compare to the $N$-body barred galaxies. Their $J$
band images, on which we have superposed some characteristic isophotes,
are given in Fig.~\ref {fig:N357_799}. Images and
results for the other galaxies in the sample can be found in
\citet{Gadotti.ACBSR07}. 

\begin{figure*}
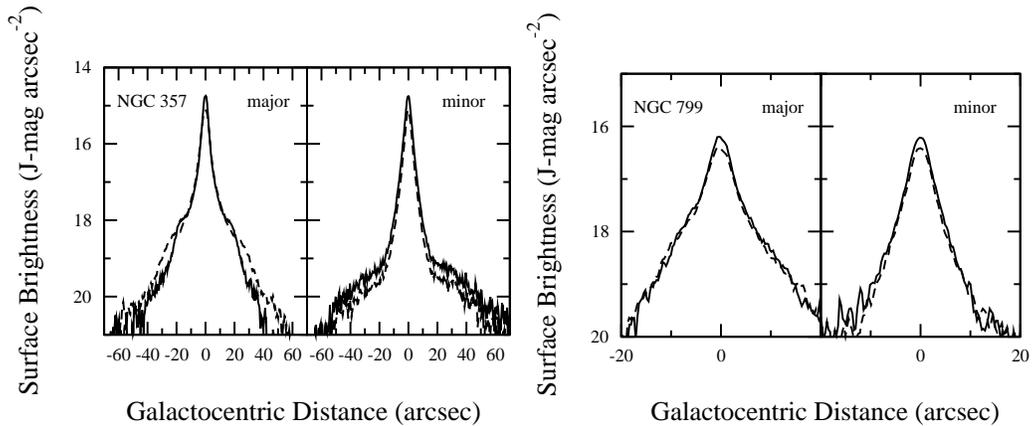

\begin{center}
  \includegraphics[clip=true,width=0.85\columnwidth]{barprof357.eps}
  \includegraphics[clip=true,width=0.85\columnwidth]{barprof799.eps}
\end{center}
  \caption{$J$ band projected surface brightness profiles along the bar
    major and minor axes. The results for the deprojected images are
    given with a dashed line.}
  \label{fig:brightnessprof}
\end{figure*}

We used the IRAF task ELLIPSE to fit ellipses to the isophotes
(Fig.~\ref{fig:ellipses}), both as 
viewed on the sky and deprojected. For each isophote, this
provides us with five very useful coefficients: the mean intensity
along the ellipse and the $a_2$, $b_2$, $a_4$ and $b_4$ coefficients
of the sin$m\theta$ and cos$m\theta$ components for $m$ = 2 and 4. In the
outer parts, the $m$ = 2 components give us information on the
position and inclination angle of the galaxy as projected on the sky
and the $m$ = 4 components are very small. At radii where the bar is
the dominant component, the $m$ = 2 components provide the bar axial
ratio and orientation and the $m$ = 4 the departure of its shape from a pure
ellipse. All this information is given in Fig.~\ref{fig:ellipsres}. 

We note that the radial profiles have quite different characteristics
for the two galaxies. The biggest difference concerns the ellipticity
radial profile. In NGC 357 this quantity rises gradually with
distance from the center, reaches a maximum and then drops very
steeply. On the other hand, in NGC 799 the gradual rise and maximum
are followed by a very gradual decrease. This difference was also seen
in the simulations, where MH type models show an abrupt drop which
lacks from MD types \citep[][ and previous
  section]{Atha.Misiriotis02}. The maximum value of the ellipticity
differs considerably between the two galaxies. For the deprojected
images, it reaches about 0.6 for NGC 357, compared to 0.4 for NGC
799. Again, this difference is characteristic of MH and MD $N$-body galaxies 
\citep[][ and previous section]{Atha.Misiriotis02}.

Note also that the $b_4$ coefficient is much larger for NGC 357 than
for NGC 799, the two maxima being of the order of 0.1 and 0.01, respectively. 
This again is a characteristic difference between MH and MD types.

We also made radial surface brightness profiles along the bar major
and minor axes and give the result in
Fig.~\ref{fig:brightnessprof}. For NGC 357 there are considerable 
differences between the profiles along these two directions. Along the
major axis the bar 
contributes a flat section to the profile, followed by a very steep
drop where it ends and joins the disc. This is absent on the minor
axis radial profile. On the other hand NGC 799 shows no flat component
corresponding to the bar and the profiles along the bar major and
minor axes are very similar. Again, this behaviour is exactly what is
expected from MH and MD $N$-body bars \citep[][ and previous
  section]{Atha.Misiriotis02}.
      
\section{Conclusions}

We can thus come to the conclusion that $N$-body simulations reproduce
well the properties of real bars and that NGC 357 is an MH type bar,
while NGC 799 is an MD type. This means that a considerable amount of
angular momentum was emitted by the near-resonant material in the bar
region of NGC 357 and absorbed by near-resonant material in its outer
disc and mainly its halo. On the other hand, much less angular
momentum must have been exchanged in NGC 799. 

\section*{Acknowledgments}
It is a pleasure for EA and AB to thank the organisers for inviting them to this
interesting meeting. 
We also thank ECOS and ANUIES for financing the exchange project
M04U01, the Agence Nationalle pour la Recherche for grant
ANR-06-BLAN-0172 and
FAPESP for grants 03/07099-0 and 00/06695-0. DAG, EA and AB would
like to thank INAOE for their kind hospitality, both in the Tonanzintla
and the Cananea sites. DAG thanks the CNRS for a 6 month
poste rouge during which the data were analysed and the project started. 
DAG is supported by the Deutsche Forschungsgemeinschaft priority program 1177 (``Witnesses of Cosmic
History: Formation and evolution of galaxies, black holes and their environment''), and the Max Planck
Society. CANICA was developed under CONACYT project
G28586E (PI: L. Carrasco). Figures 2, 3 and 4 are part of Figures 2
and 3 of Gadotti et al. (2007), which is published by Blackwell publishers
in `Monthly Notices of the Royal Astronomical Society' and is
available at www.blackwell-synergy.com.

\end{document}